\documentclass[pra,twocolumn,tightenlines,showpacs,nofootinbib]{revtex4}
\usepackage{bm,dcolumn,amsmath,graphicx,amsfonts,amssymb}

\begin{document}
\title{Calculation of nuclear-spin-dependent parity nonconservation in
  s-d transitions of Ba$^+$, Yb$^+$ and Ra$^+$ ions.}

\author{V. A. Dzuba and V. V. Flambaum}
\affiliation{School of Physics, University of New South Wales,
Sydney, NSW 2052, Australia}

\date{ \today }

\begin{abstract}
We use correlation potential and many-body perturbation theory
techniques to calculate spin-independent and nuclear spin-dependent parts of
the parity nonconserving amplitudes of the transitions between the
$6s_{1/2}$ ground state and the $5d_{3/2}$ excited state of Ba$^+$ and
Yb$^+$ and between the $7s_{1/2}$ ground state and the $6d_{3/2}$
excited state of Ra$^+$. The results are presented in a form
convenient for extracting of the constants of nuclear-spin-dependent
interaction (such as, e.g., anapole moment) from the measurements. 
\end{abstract}
\pacs{11.30.Er, 31.15.A-}
\maketitle

\section{Introduction}

The study of the parity nonconservation (PNC) in atoms is a
low-energy, relatively inexpensive alternative to high-energy search
for new physics beyond the standard model (see, e.g. \cite{Khrip}). 
The most significant recent achievement on this
path is the very precise measurements of the PNC in
cesium~\cite{Wood}. The cesium PNC
experiment together with its interpretation~\cite{Breit,QED,Cs-cor} in
terms of nuclear weak 
charge provides the best current atomic test of the standard model
(see also review~\cite{Ginges}). It is also the only measurement
of the nuclear {\em anapole moment} which is produced by the PNC nuclear
forces \cite{anapole}. The extraction of the weak
nuclear charge from the PNC measurements relies on atomic
calculations. Cesium atom has the simplest electron structure among
all heavy atoms which were used or considered for the PNC
measurements. Still it took considerable efforts of several groups of
theorists to bring the accuracy of the calculations in line with the
accuracy of measurements and provide reliable interpretation of the
measurements in terms of the standard model and possible new physics
beyond it~\cite{Breit,QED,Cs-cor}.
It is widely believed now that it would be hard to compete with 
cesium experiment in terms of accuracy of interpretation of the PNC
measurements. Therefore, the study of PNC in atoms is mostly focused
now in two directions: (i) the measurements of the PNC ratio for a
chain of isotopes which was first proposed in Ref.~\cite{DFK86}, and
(ii) the measurements of the nuclear-spin-dependent 
PNC, like e.g. the contribution from nuclear anapole moment (see,
e.g. reviews \cite{Ginges,DF10}).
The study of the PNC for a chain of isotopes
does not require atomic calculations and can deliver useful
information about either neutron distribution or new physics beyond
standard model (see, e.g.~\cite{Fortson,DP02,BDF09}). The measurements
of anapole moment does require atomic calculations but high accuracy
is not critical here.

Ba$^+$, Yb$^+$ and Ra$^+$ ions considered in present paper are good
candidates for both types of the experimental studies. 
Ba and Yb both have seven stable isotopes with large difference in neutron
numbers $\Delta N_{max} = 8$. Radium has several long-living isotopes.
There are two stable isotopes for each of the Ba and Yb atoms
($^{135}$Ba, $^{139}$Ba, $^{171}$Yb and $^{173}$Yb) which have
non-zero nuclear spin. There are also isotopes of Ra with non-zero
nuclear spin ($^{223}$Ra, $^{225}$Ra, $^{229}$Ra). In all cases
nuclear spin is provided by valence neutron. 
This is especially interesting since 
it allows one to measure the strength of the neutron-nucleus PNC potential
\cite{anapole} (the anapole moment has been measured only for the $^{133}$Cs
nucleus which has valence proton). 

Finally, Ba$^+$ and Ra$^+$ ions have electron structure similar to
those of cesium atom. This means that the accuracy of the
interpretation of the PNC measurements can be on the same level as for
cesium. Moreover, it can be further improved with the use of the
experimental data~\cite{DFG01}. 

The use of Ba$^+$ in the PNC measurements was first suggested by
Fortson~\cite{F93}. The work is in progress at Seattle (see,
e.g. \cite{Sherman05,Sherman}) but no PNC results have been reported
yet. Similar approach is now considered for the measurements of PNC in
Ra$^+$ ion at KVI~\cite{Wansbeek,Versolato}. It is important that in Ra$^+$ the
PNC effects are about 20 times larger than in Ba$^+$. There are plans to
measure PNC in Yb$^+$ at Los Alamos~\cite{Torgerson}. Note that the
PNC measurements for neutral ytterbium are in progress at Berkeley and
first PNC results were recently reported~\cite{BudkerYbPNC}. The PNC
measurements for the Yb$^+$ ion would provide an important consistency
test for the measurements and their interpretation.

Calculations of the spin-independent PNC amplitude for Ba$^+$ and
Ra$^+$ were performed in our early work~\cite{DFG01} and in
\cite{Sahoo06}. Calculations for 
Ra$^+$ were later performed in \cite{Wansbeek} and \cite{Pal09}. The
only calculation of the spin-dependent PNC in Ra$^+$ was recently
reported by Sahoo {\em et al}~\cite{Sahoo11}. To the best of our
knowledge, no PNC calculations for Yb$^+$ have been published so far.

In present paper we calculate both spin-independent and spin-dependent
PNC amplitudes simultaneously using the same procedure and the same wave
functions. In this approach the relative sign of the amplitudes is
fixed. This allows for unambiguous determination of the sign of the
spin-dependent contribution. The constant of the spin-dependent interaction
can be expressed via the ratio of the two amplitudes. This
brings an extra advantage of more accurate interpretation of the
measurements. The accuracy of the calculations for the ratio of the PNC
amplitudes is usually higher than that for each of the amplitudes. This is
because the amplitudes are often very similar in structure and most of the
theoretical uncertainty cancels out in the ratio.

Since we focus on the calculation of the nuclear-spin-dependent PNC
amplitudes where high accuracy of calculations is not needed, we don't
include some small corrections, like some classes of diagrams for
higher-order correlations, Breit and quantum electrodynamic (QED)
corrections, etc. Instead, we make sure that all leading contributions
are included exactly the same way for both spin-independent and
spin-dependent PNC amplitudes which is important for the cancelation
of the uncertainty in the ratio.

\section{Theory}

\label{theory}
Hamiltonian describing parity-nonconserving electron-nuclear
interaction can be written as a sum of spin-independent (SI) and
spin-dependent (SD) parts (we use atomic units: $\hbar = |e| = m_e = 1$):
%------------------------------------------------------------------
\begin{eqnarray}
     H_{\rm PNC} &=& H_{\rm SI} + H_{\rm SD} \nonumber \\
      &=& \frac{G_F}{\sqrt{2}}                             %  (1)
     \Bigl(-\frac{Q_W}{2} \gamma_5 + \frac{\varkappa}{I}
     {\bm \alpha} {\bm I} \Bigr) \rho({\bm r}),
\label{e1}
\end{eqnarray}
%------------------------------------------------------------------
where  $G_F \approx 2.2225 \times 10^{-14}$ a.u. is the Fermi constant of
the weak interaction, $Q_W$ is the nuclear weak charge,
$\bm\alpha=\left(
\begin{array}
[c]{cc}%
0 & \bm\sigma\\
\bm\sigma & 0
\end{array}
\right)$ and $\gamma_5$ are the Dirac matrices, $\bm I$ is the
nuclear spin, and $\rho({\bf r})$ is the nuclear density normalized to 1.
The strength of the spin-dependent PNC interaction is proportional to
the dimensionless constant $\varkappa$ which is to be found from the
measurements. There are three major contributions to
$\varkappa$ arising from (i) electromagnetic interaction of atomic
electrons with nuclear {\em anapole moment}~\cite{FKh85}, (ii)
electron-nucleus spin-dependent weak interaction, and (iii) combined effect of
spin-independent weak interaction and magnetic hyperfine interaction~\cite{Novikov}
(see, also review~\cite{Ginges}). In this work we do not distinguish
between different contributions to $\varkappa$ and present the results
in terms of total $\varkappa$ which is the sum of all possible
contributions. 

Within the standard model
the weak nuclear charge $Q_W$ is given by~\cite{PDG}
\begin{equation}
%Q_W = -N + Z\,(1-4\,\sin^2\theta_W) .
Q_W \approx -0.9877N + 0.0716Z.
\label{eq:qw}
\end{equation}
Here $N$ is the number of neutrons, $Z$ is the number of protons.
%and $\theta_W$ is the Weinberg angle.

The PNC amplitude of an electric dipole transition between states of
the same parity $|i\rangle$ and $|f \rangle$ is equal to:
\begin{eqnarray}
   E1^{PNC}_{fi}  &=&  \sum_{n} \left[
\frac{\langle f | {\bm d} | n  \rangle
      \langle n | H_{\rm PNC} | i \rangle}{E_i - E_n}\right.
\nonumber \\
      &+&
\left.\frac{\langle f | H_{\rm PNC} | n  \rangle
      \langle n | d_q | i \rangle}{E_f - E_n} \right],
\label{eq:e2}
\end{eqnarray}
where ${\bm d} = -e\sum_i {\bm r_i}$ is the electric dipole operator,
  $|a \rangle \equiv |J_a F_a M_a \rangle$ and ${\bm F} = {\bm I}
+ {\bm J}$ is the total angular momentum. 

Applying the Wigner-Eckart theorem we can express the amplitudes via
reduced matrix elements
\begin{eqnarray}
  E1^{PNC}_{fi} &=&
      (-1)^{F_f-M_f} \left( \begin{array}{ccc}
                           F_f & 1 & F_i  \\
                          -M_f & q & M_i   \\
                           \end{array} \right) \nonumber \\
   &\times& \langle J_f F_f || d_{\rm PNC} || J_i F_i \rangle .
\end{eqnarray}
Detailed expressions for the reduced matrix elements of the SI and
SD PNC amplitudes can be found e.g. in Refs.~\cite{Porsev01} and
\cite{JSS03}. For the SI amplitude we have
\begin{eqnarray}
&&\langle J_f,F_f || d_{\rm SI} || J_i,F_i \rangle =
(-1)^{I+F_i+J_f+1}\nonumber \\ 
&& \times \sqrt{(2F_f+1)(2F_i+1)} 
\left\{ \begin{array}{ccc} J_i & J_f & 1 \\
                          F_f & F_i & I \\ 
                    \end{array} \right\}  \label{eq:si0}\\
&&  \times \sum_{n} \left[
\frac{\langle J_f || {\bm d} || n,J_n  \rangle
      \langle n,J_n || H_{\rm SI} || J_i \rangle}{E_i - E_n}\right.  \nonumber \\
&& + \left.\frac{\langle J_f || H_{\rm SI} || n,J_n  \rangle
      \langle n,J_n || {\bm d} || J_i \rangle}{E_f - E_n} \right]. \nonumber 
%&& \equiv c(F_f,J_f,F_i,J_i) E^{\prime}_{fi}. \nonumber
\end{eqnarray}
%Here $c(F_f,J_f,F_i,J_i)$ is the angular coefficient and the sum over $n$,
%$E^{\prime}_{fi}$ does not depend on $F_f$ or $F_i$:
%\begin{eqnarray}
%  E^{\prime}  &=&  \sum_{n} \left[
%\frac{\langle J_f || {\bm d} || n,J_n  \rangle
%      \langle n,J_n || H_{\rm SI} || J_i \rangle}{E_i -
%      E_n}\right. \label{eq:si} \\ 
%&+&  \left.\frac{\langle J_f || H_{\rm SI} || n,J_n  \rangle
%      \langle n,J_n || {\bm d} || J_i \rangle}{E_f - E_n} \right]. \nonumber 
%\end{eqnarray}

For the SD PNC amplitude we have
\begin{eqnarray}
    && \langle J_f,F_f || d_{\rm SD} || J_i,F_i \rangle =
    \frac{G_F}{\sqrt{2}} \varkappa \nonumber \\
     &&\times  \sqrt{(I+1)(2I+1)(2F_i+1)(2F_f+1)/I}  \nonumber \\
    &&\times
     \sum_{n} \left[ (-1)^{J_f - J_i}
     \left\{ \begin{array}{ccc}
     J_n  &  J_i  &   1    \\
      I   &   I   &  F_i   \\                                  
     \end{array} \right\}
     \left\{ \begin{array}{ccc}
      J_n  &  J_f  &  1   \\
      F_f  &  F_i  &  I   \\
     \end{array} \right\} \right. \nonumber \\
  &&\times \frac{ \langle J_f || {\bm d} || n, J_n \rangle
     \langle n, J_n || {\bm \alpha}\rho || J_i \rangle }{E_n -
     E_i} \label{eq:dsd}  \\
  &&+
     (-1)^{F_f - F_i}
     \left\{ \begin{array}{ccc}
     J_n  &  J_f  &   1    \\
      I   &   I   &  F_f   \\
     \end{array} \right\}
     \left\{ \begin{array}{ccc}
     J_n  &  J_i  &  1   \\
     F_i  &  F_f  &  I   \\
     \end{array} \right\} \nonumber \\
 &&\times
     \left. \frac{\langle J_f || {\bm \alpha}\rho ||n,J_n \rangle
            \langle n,J_n || {\bm d} ||J_i \rangle}{E_n - E_f}  \right].
\nonumber
\end{eqnarray}
For the case of the $5d - 6s$ transitions considered in present paper
(or $6d-7s$ in the case of Ra$^+$) it is convenient to break 
expression (\ref{eq:dsd}) into four parts:
\begin{equation}
  \langle 5d_{3/2},F_f || d_{\rm SD} || 6s,F_i \rangle = S_1+S_2+S_3+S_4,
\label{eq:ssss}
\end{equation}
where
\begin{eqnarray}
  &&S_1 = c_1(F_f,F_i) \label{eq:s1} \\
 &&\times \sum_n \frac{ \langle 5d_{3/2} || {\bm d} || np_{1/2} \rangle
     \langle np_{1/2} || {\bm \alpha}\rho || 6s \rangle }{E_{np_{1/2}} -
     E_{6s}} ,  \nonumber \\
  &&S_2 = c_2(F_f,F_i) \label{eq:s2} \\
  &&\times \sum_n \frac{ \langle 5d_{3/2} || {\bm d} || np_{3/2} \rangle
     \langle np_{3/2} || {\bm \alpha}\rho || 6s \rangle }{E_{np_{3/2}} -
     E_{6s}} , \nonumber  \\
   &&S_3 = c_3(F_f,F_i) \label{eq:s3} \\ 
   &&\times \sum_n \frac{\langle 5d_{3/2} || {\bm \alpha}\rho ||np_{1/2} \rangle
            \langle np_{1/2} || {\bm d} ||6s \rangle}{E_{np_{1/2}} -
            E_{5d_{3/2}}}, \nonumber \\ 
   &&S_4 = c_4(F_f,F_i) \label{eq:s4} \\ 
   && \times \sum_n \frac{\langle 5d_{3/2} || {\bm \alpha}\rho
     ||np_{3/2} \rangle \langle np_{3/2} || {\bm d} ||6s
     \rangle}{E_{np_{3/2}} - E_{5d_{3/2}}}. \nonumber
\end{eqnarray}
Here $c_m(F_f,F_i)$ ($m=1,2,3,4$) are coefficients which can be
reconstructed using (\ref{eq:dsd}). The terms $S_1, S_2, S_3, S_4$
differ by the order of the operators $\bm d$ and ${\bm \alpha}\rho$
and by the states in the summation which are either $np_{1/2}$ or
$np_{3/2}$ states. To know the relative values of these terms is
important for the analysis of the accuracy of
the calculations.

\section{Calculations}

\label{calculations}
To perform the calculations we follow an {\em ab initio} approach
which uses the correlation potential method~\cite{CPM} and the
technique to include higher-order correlations developed in
Refs.~\cite{DFS89a,DFS89b,DFKS}.

Calculations start from the relativistic Hartree-Fock (RHF) method
in the $V^{N-1}$ approximation. This means that the initial RHF
procedure is done for a closed-shell atomic core with the valence
electron removed. After that, the states of the external electron are
calculated in the field of the frozen core. Correlations are
included by means of the correlation potential method~\cite{CPM}.
For Ba$^+$ and Ra$^+$ we use the all-order correlation potential $\hat
\Sigma^{(\infty)}$ which includes two classes of the higher-order
terms: screening of the Coulomb interaction and hole-particle
interaction (see, e.g.~\cite{DFS89a} for details). For Yb$^+$ we use
the second-order correlation potential $\hat \Sigma^{(2)}$. The reason for
different approaches is due to different electron structures of the
ions. The all-order technique developed in \cite{DFS89a,DFS89b,DFKS}
works very well for alkali atoms and similar ions in which the valence
electron is far from the atomic core and
higher-order correlations are dominated by screening of the core-valence
residual Coulomb interaction by the core electrons. For atoms and ions similar to
Yb$^+$, in which an external electron is close to the core and
strongly interacts with its electrons, a different higher-order
effect described by the {\em ladder diagrams}~\cite{ladder} becomes
important. The applicability of the technique of
Ref.~\cite{ladder} to Yb$^+$ needs further investigation. Meanwhile,
the use of the second-order $\hat \Sigma^{(2)}$ leads to sufficiently good
results. Note that an external electron in Ba$^+$ and Ra$^+$ ions is
also closer to atomic core than in neutral alkali atoms Cs and Fr.
This means that inclusion of ladder diagrams might be a way to improve
the accuracy of calculations for the ions as well. This question also
needs further investigation.
  
To calculate $\hat \Sigma$ ($\hat \Sigma^{(\infty)}$ or $\hat
\Sigma^{(2)}$)we need a complete set of the single-electron 
orbitals. We use the B-spline technique~\cite{Bspline} to
construct the basis. The orbitals are built as linear combinations of
50 B-splines of order 9 in a cavity of radius 40$a_B$.
The coefficients are chosen from the condition that the
orbitals are the eigenstates of the RHF Hamiltonian $\hat H_0$ of the
closed-shell core. The second-order operator $\hat \Sigma^{(2)}$ is
calculated via direct summation over B-spline basis states.
The all-order $\hat \Sigma^{(\infty)}$ is
calculated with the technique which combines solving equations for
the Green functions (for the direct diagram) with the summation over
complete set of states (exchange diagram)~\cite{DFS89a}.

The correlation potential $\hat \Sigma$ is then used to build a new
set of single-electron states, the so-called Brueckner orbitals.
This set is to be used in the summation in equations (\ref{eq:si0}),
and (\ref{eq:dsd}). Here again we use the B-spline
technique to build the basis. The procedure is very similar to
constructing of the RHF B-spline basis. The only difference is that
new orbitals are now the eigenstates of the $\hat H_0 + \hat \Sigma$
Hamiltonian.

\begin{table}
\caption{Ionization energies of lowest $s,p$ and $d$ states of Ba$^+$,
  Yb$^+$ and Ra$^+$  in different approximations (cm$^{-1}$).}
\label{t:en}
\begin{ruledtabular}
\begin{tabular}{ccccc}
Ion & State & RHF & Brueckner & Experiment\cite{web} \\
\hline
Ba$^+$ & $6s_{1/2}$ & 75340 & 80815 & 80687 \\
       & $6p_{1/2}$ & 57266 & 60571 & 60425 \\
       & $6p_{3/2}$ & 55873 & 58848 & 58735 \\
       & $5d_{3/2}$ & 68139 & 76318 & 75813 \\
\hline                          
Yb$^+$ & $6s_{1/2}$  & 90789 & 99477 & 98207 \\
       & $6p_{1/2}$  & 66087 & 70728 & 71145 \\
       & $6p_{3/2}$  & 63276 & 67101 & 67815 \\
       & $5d_{3/2}$  & 66517 & 75551 & 75246 \\
\hline
Ra$^+$ & $7s_{1/2}$ & 75898 & 82032 & 81842 \\
       & $7p_{1/2}$ & 56878 & 60715 & 60491 \\
       & $7p_{3/2}$ & 52906 & 55753 & 55633 \\
       & $6d_{3/2}$ & 62356 & 70091 & 69758 \\
\end{tabular}
\end{ruledtabular}
\end{table}

\begin{table}
\caption{Rescaling factors for the correlation potential $\hat \Sigma$.}
\label{t:f}
\begin{ruledtabular}
\begin{tabular}{ccccc}
Ion & $s_{1/2}$ & $p_{1/2}$ & $p_{3/2}$ & $d_{3/2}$ \\
\hline
Ba$^+$ & 0.978 & 0.960 & 0.964 & 0.941 \\
Yb$^+$ & 0.862 & 1.081 & 1.170 & 0.968 \\
Ra$^+$ & 0.970 & 0.946 & 0.960 & 0.959 \\
\end{tabular}
\end{ruledtabular}
\end{table}

Brueckner orbitals which correspond to the lowest valence states are
good approximations to the real physical states. Their quality can
be tested by comparing experimental and theoretical energies.
The energies of the lowest states of Ba$^+$, Yb$^+$ and Ra$^+$ in RHF
and Brueckner approximations are presented in Table~\ref{t:en}.
One can see that inclusion of the correlations leads to significant
improvement of the accuracy in all cases. The deviation of the theory
from experiment is just fraction of a per cent in the case of Ba$^+$
and Ra$^+$ where an all-order $\hat \Sigma^{(\infty)}$ is used and
does not exceed 1.3\% for Yb$^+$ where the second-order $\hat
\Sigma^{(2)}$ is used.  

The quality of the Brueckner orbitals can be further improved by rescaling the
correlation potential $\hat \Sigma$ to fit the experimental energies
exactly. We do this by replacing the $\hat H_0 + \hat \Sigma$ with
the $\hat H_0 + \lambda \hat \Sigma$ Hamiltonian in which the
rescaling parameter $\lambda$ is chosen for each partial wave to fit
the energy of the first valence state. The values of $\lambda$ are
presented in Table~\ref{t:f}.
Note that these values are very close to unity.
This means that even without rescaling the accuracy is
good and only a small adjustment of the value of $\hat \Sigma$ is
needed. Note also that since the rescaling procedure affects not
only energies but also the wave functions, it usually leads to
improved values of the matrix elements of external fields. In fact,
this is a semi-empirical method to include omitted higher-order
correlation corrections.

Matrix elements of the $H_{\rm SI}$, $H_{\rm SD}$ and electric dipole
operators are found by means of the time-dependent Hartree-Fock (TDHF)
method~\cite{CPM,DFS84} extended to Brueckner orbitals. This method 
incorporates to the well-known random-phase approximation (RPA)
diagrams including exchange. In the TDHF method, the 
single-electron wave functions are presented in the form $\psi =
\psi_0 + \delta \psi$, where $\psi_0$ is the unperturbed wave
function. It is an eigenstate of the RHF Hamiltonian $\hat H_0$: 
$(\hat H_0 -\epsilon_0)\psi_0 = 0$.  $\delta \psi$ is the correction
due to external field. It can be found be solving the TDHF equation
\begin{equation}
    (\hat H_0 -\epsilon_0)\delta \psi = -\delta\epsilon \psi_0 - \hat F \psi_0 -
  \delta \hat V^{N-1} \psi_0,
  \label{TDHF}
\end{equation}
where $\delta\epsilon$ is the correction to the energy due to external field
($\delta\epsilon\equiv 0$ for all above mentioned  operators but it is not
zero for the hyperfine interaction which we will need for the analysis
of accuracy), $\hat F$ is the operator of the external field, and
$\delta \hat V^{N-1}$ is the correction to the self-consistent
potential of the core due to external field. 

The TDHF equations are solved self-consistently for all states in the
core. Then the matrix elements between any (core or valence) states
$n$ and $m$ are given by 
\begin{equation}
    \langle \psi_n | \hat F + \delta \hat V^{N-1} | \psi_m \rangle.
    \label{mel}
\end{equation}

The best results are achieved when $\psi_n$ and $\psi_m$ are the Brueckner
orbitals computed with rescaled correlation potential $\hat \Sigma$.

We use equation (\ref{mel}) for all weak and electric dipole matrix
elements in evaluating the SI and SD PNC amplitudes (\ref{eq:si0}) and
(\ref{eq:dsd}).

\section{Accuracy of calculations}

\label{accuracy}

\begin{table}
\caption{Electric dipole matrix elements. Comparison of present
  calculations with experiment or most complete other calculations.} 
\label{t:E1}
\begin{ruledtabular}
\begin{tabular}{cccl}
Ion & Transition & This work & Other \\
\hline
Ba$^+$ & $6s_{1/2}-6p_{1/2}$ & 3.32 & 3.36(4)\footnotemark[1] \\
       & $6s_{1/2}-6p_{3/2}$ & 4.69 & 4.55(10)\footnotemark[1] \\
       & $5d_{3/2}-6p_{1/2}$ & 3.06 & 3.14(8)\footnotemark[2] \\
       & $5d_{3/2}-6p_{3/2}$ & 1.34 & 1.54(19)\footnotemark[1] \\
\hline
Yb$^+$ & $6s_{1/2}-6p_{1/2}$ & 2.72 & 2.471(3)\footnotemark[3] \\
       & $6s_{1/2}-6p_{3/2}$ & 3.84 & 3.36(2)\footnotemark[4] \\
       & $5d_{3/2}-6p_{1/2}$ & 3.09 & 2.97(4)\footnotemark[3] \\
       & $5d_{3/2}-6p_{3/2}$ & 1.36 & 1.31\footnotemark[5] \\
\hline
Ra$^+$ & $7s_{1/2}-7p_{1/2}$ & 3.24 & 3.254\footnotemark[6] \\
       & $7s_{1/2}-7p_{3/2}$ & 4.49 & 4.511\footnotemark[6] \\
       & $6d_{3/2}-7p_{1/2}$ & 3.56 & 3.566\footnotemark[6] \\
       & $6d_{3/2}-7p_{3/2}$ & 1.51 & 1.512\footnotemark[6] \\

\end{tabular}
\footnotemark[1]{Experiment, Ref.~\cite{Davidson}.}
\footnotemark[2]{Experiment, Ref.~\cite{Sherman}.}
\footnotemark[3]{Experiment, Ref.~\cite{Olmschenk1,Olmschenk2}.}
\footnotemark[4]{Experiment, Ref.~\cite{Pinnington}.}
\footnotemark[5]{Theory, Ref.~\cite{SS09}.}
\footnotemark[6]{Theory, Ref.~\cite{Pal09}.}
\end{ruledtabular}
\end{table}

\begin{table}
\caption{Magnetic dipole hyperfine constants $A$ (MHz). Comparison of
  present calculations with experiment.}
\label{t:hfs}
\begin{ruledtabular}
\begin{tabular}{ccrl}
Ion & State & This work & Experiment \\
\hline
$^{135}$Ba$^+$ & $6s_{1/2}$ & 3671 & 3593.3(2.2)\footnotemark[1] \\
              & $6p_{1/2}$ &  668 & 664.6(0.3)\footnotemark[2] \\
              & $6p_{3/2}$ &  131 & 113.0(0.1)\footnotemark[2] \\
              & $5d_{3/2}$ &  161 & 169.5892(9)\footnotemark[3] \\
\hline
$^{171}$Yb$^+$ & $6s_{1/2}$ & 13217 & 12645(2)\footnotemark[4] \\
              & $6p_{1/2}$ &  2533 & 2104.9(1.3)\footnotemark[4] \\
              & $6p_{3/2}$ &  388 & 877(20)\footnotemark[5] \\
              & $5d_{3/2}$ &  291 & 430(43)\footnotemark[6] \\
\hline
$^{223}$Ra$^+$ & $7s_{1/2}$ & 3537 & 3404(2)\footnotemark[7] \\
              & $7p_{1/2}$ &  679 &  667(2)\footnotemark[7] \\
              & $7p_{3/2}$ &  69.8 & 56.5(8)\footnotemark[7]\\
              & $6d_{3/2}$ &  57.8 & 77.6(8)\footnotemark[8]\\

\end{tabular}
\footnotemark[1]{Ref.~\cite{Wendt84}.}
\footnotemark[2]{Ref.~\cite{Villemoes}.}
\footnotemark[3]{Ref.~\cite{Silverans}.}
\footnotemark[4]{Ref.~\cite{MP94}.}
\footnotemark[5]{Ref.~\cite{Berends}.}
\footnotemark[6]{Ref.~\cite{Engelke}.}
\footnotemark[7]{Ref.~\cite{Wendt87,Neu}.}
\footnotemark[8]{Rescaled from $^{213}$Ra~\cite{Versolato} using
  magnetic moments from \cite{Arnold}.}
\end{ruledtabular}
\end{table}

The accuracy of the results obtained via direct summation over
physical states with the use of expressions like (\ref{eq:e2}) is
determined by the accuracy for the energies, electric dipole and weak
matrix elements. We start from the notion that for the PNC amplitudes
considered in present work the summation over intermediate $p$-states
is strongly dominated by the $6p_{1/2}$ and $6p_{3/2}$ states for Ba$^+$
and Yb$^+$ and by $7p_{1/2}$ and $7p_{3/2}$ states for
Ra$^+$. Corresponding contributions constitute 70 to 90\% of the total
PNC amplitude. Therefore, it is sufficient to compare with experiment
energies and matrix elements involving these $p$-states. The energies
and electric dipole matrix elements can be directly compared with
experiment while standard practice of comparing experimental and
theoretical hyperfine structure can be used to test the
accuracy of the weak matrix elements. 

To improve the accuracy for the amplitudes the energies of the $6s$,
$6p_{1/2}$,  $6p_{3/2}$ and  $5d_{3/2}$ states ($7s$, $7p_{1/2}$,
$7p_{3/2}$ and  $6d_{3/2}$ for Ra$^+$) are fitted exactly in our
calculations using rescaling of the correlation potential $\hat
\Sigma$ as it has been described in previous section.  

Calculated and experimental E1-transition amplitudes are presented in
Table~\ref{t:E1}. Note that we need comparison with experiment only
for estimation of the accuracy of our calculations. Therefore,
a comprehensive review of the experimental and theoretical data
available for the ions goes beyond the scope of present work. 
We only compare our results with the most accurate experimental data
or with the most complete other calculations where the experimental data
are not available. Good reviews of the electric dipole transition data
in Ba$^+$ and Yb$^+$ can be found in Ref.~\cite{Sherman} and
\cite{Olmschenk2}.  

The data in Table~\ref{t:E1} shows good agreement between theory and
experiment for most of the amplitudes, although the accuracy for the
amplitudes involving the $p_{3/2}$ states is lower than that for the
$p_{1/2}$ states.

Table~\ref{t:hfs} shows theoretical and experimental data on the
hyperfine structure constants of the low states of Ba$^+$, Yb$^+$ and
Ra$^+$. Here again we only compare our calculations with the most
accurate experimental data. A review of the available experimental and
theoretical data for Ba$^+$ can be found in Ref.~\cite{Yu09}.
The data in Table~\ref{t:hfs} shows several trends: (i) the accuracy
is good for $s_{1/2}$ and $p_{1/2}$ states, especially in the cases of
Ba$^+$ and Ra$^+$, (ii) the accuracy for Yb$^+$ is lower than that for
Ba$^+$  and Ra$^+$, (iii) the accuracy for $p_{3/2}$ and $d_{3/2}$
states is lower than that for the $s_{1/2}$ and $p_{1/2}$ states. 
The largest discrepancy is for the hfs
of the $6_{3/2}$ state of Yb$^+$ where theory and experiment differ 
almost three times. Note that the most complete calculations of
Ref.~\cite{SS09} give the result which is close to our theoretical
value rather than to the experiment. In principle, the discrepancy
can be explained by configuration mixing involving configurations with
excitations from the $4f$ subshell. Neither our present calculations
nor those of Ref.~\cite{SS09} include this mixing explicitly. The
configuration interaction calculations based on technique developed in
Ref.~\cite{DF08a,DF08b} which treats Yb$^+$ as a system with fifteen
valence electrons show that the hfs of the $6p_{3/2}$ state is indeed
very sensitive to the configuration mixing. One can find such mixing
which reproduces the experimental hfs exactly while the accuracy for
the energy and for the $g$-factor of the $6p_{3/2}$ state is also
good. However, the results are inconclusive due to strong instability
of the hfs of the $6p_{3/2}$ state. We can only say that the
configuration mixing can explain current experimental value of the hfs
of $6p_{3/2}$ state but we cannot prove that this explanation is
correct. Since the disagreement between 
theory and experiment for the hfs of the $6p_{3/2}$ state of Yb$^+$ is
the main factor contributing to the uncertainty of the calculations
for Yb$^+$, it would be useful to remeasure the hfs of this state.

The fact that the accuracy for the $p_{1/2}$ and $p_{3/2}$ states is
different complicates the analysis of the accuracy for the PNC
amplitudes. There is cancelation between terms containing matrix
elements with the $p_{1/2}$ and $p_{3/2}$ states. In the end of
section \ref{theory} we introduced the notations $S_1, S_2, S_3$ and
$S_4$ for these terms (see
Eqs.(\ref{eq:s1},\ref{eq:s2},\ref{eq:s3},\ref{eq:s4})). The terms
involving the $p_{1/2}$ states are 
$S_1$ and $S_3$, the terms with the $p_{3/2}$ states are $S_2$ and
$S_4$. Table~\ref{t:ssss} shows the $S_1, S_2, S_3, S_4$ contributions
to the reduced matrix elements of the nuclear-spin-dependent PNC interaction
in some hfs components of the transitions in Ba$^+$, of Yb$^+$ and of
Ra$^+$. One can see that the $S_2$ term is usually small while the
$S_4$ term is not small. For example, for Yb$^+$ the contribution of
the $S_4$ term is more than a half of the total sum. It is clear that
the accuracy of the calculations in this case will be mostly
determined by the accuracy of the $S_4$ term.

To analyse the accuracy of the PNC calculations we need a procedure
which takes into account the deviation of the experimental and
theoretical data for the electric dipole matrix elements and for the
hyperfine structure as well as the effect of partial cancelation
between different contributions to the PNC amplitude. We do this by
comparing the {\em   ab initio} calculations with the calculations in
which the electric dipole and weak matrix elements are rescaled to fit
the experimental data. For example, assuming that the weak matrix
elements between two states are proportional to the square root of the
hfs constants for these states we rescale them as following
\begin{equation}
\langle n |H_{\rm PNC}|m \rangle_{\rm rescaled} = 
\sqrt{\frac{A_n^{\rm exp}A_m^{\rm exp}}{A_n^{\rm th}A_m^{\rm th}}}
\langle n |H_{\rm PNC}|m \rangle.
\label{eq:rs}
\end{equation}
Here $A_n^{\rm exp}$ and $A_n^{\rm th}$ are experimental and
theoretical values of the hfs constants from Table~\ref{t:hfs}. 
This means that we perform accurate rescaling for matrix elements
involving $6p_{1/2}$ and $6p_{3/2}$ states ($7p_{1/2}$ and $7p_{3/2}$
for Ra$^+$). As it was stated above, this corresponds to 70 to 90\% of
the total PNC amplitude. We use the same rescaling for all matrix
elements involving higher $p$ states. Electric dipole matrix elements
are also rescaled to fit the experimental data for the transitions
between lowest states. The difference between PNC amplitudes obtained
in the {\em ab initio} calculations and calculations with rescaling
serves as an estimation of the uncertainty of the calculations.   

Note that the accuracy for the relative contribution of the
nuclear-spin-dependent interaction can be higher that for  
each of the amplitudes (see also Ref.~\cite{DF11}). As we will see in
the next section, this is usually the case when the $S_2$ and $S_3$
contributions are both small. This is because these terms are exactly
zero for the spin-independent PNC amplitudes. Therefore, the
spin-dependent PNC amplitudes in which the $S_2$ and $S_3$ terms are
small, are similar to the spin-independent amplitudes. They both change
under scaling at the same rate which cancels out in the ratio. 

\begin{table}
\caption{Contributions to the reduced matrix elements 
$\langle 5d_{3/2},F_1||\hat H^{\rm eff}_{\rm SDPNC}||6s_{1/2},F_2\rangle$ of
the spin-dependent parity-nonconserving s-d transitions. See text for
explanation of notations.
Units: $10^{-11} \varkappa iea_0$.} 
\label{t:ssss}
\begin{ruledtabular}
\begin{tabular}{ccc rrrrr}
Ion & $F_1$ & $F_2$ & $S_1$ & $S_2$ & $S_3$ &
$S_4$ & Sum \\
\hline
$^{135}$Ba$^+$ & 0 & 1 &  0.134 &  0.002 &  0.000 & -0.027 &  0.108 \\
              & 1 & 1 & -0.211 & -0.001 &  0.013 &  0.032 & -0.168 \\
              & 1 & 2 & -0.057 &  0.003 &  0.029 & -0.014 & -0.040 \\
              & 2 & 1 &  0.211 & -0.002 & -0.038 & -0.009 &  0.162 \\
              & 2 & 2 &  0.127 & -0.003 & -0.038 &  0.009 &  0.094 \\
              & 3 & 2 & -0.212 & -0.002 &  0.000 &  0.043 & -0.171 \\
\hline
$^{171}$Yb$^+$ & 1 & 0 &  0.780 &  0.000 & -0.306 & -0.164 &  0.310 \\
              & 1 & 1 &  0.184 & -0.008 & -0.432 &  0.116 & -0.140 \\
              & 2 & 1 & -0.411 & -0.004 &  0.000 &  0.156 & -0.259 \\
\hline
%$^{225}$Ra$^+$ & 1 & 0 &  3.536 &  0.000 & -0.475 & -0.116 &  2.944 \\
%              & 1 & 1 &  0.833 & -0.065 & -0.672 &  0.082 &  0.178 \\
%              & 2 & 1 & -1.864 & -0.029 &  0.000 &  0.110 & -1.782 \\
$^{229}$Ra$^+$ & 1 & 2 &   2.021 &   0.031 &   0.000 &  -0.119 &   1.933 \\
              & 2 & 2 &  -2.301 &  -0.005 &   0.265 &   0.084 &  -1.957 \\
              & 2 & 3 &  -0.878 &   0.044 &   0.496 &  -0.045 &  -0.384 \\
              & 3 & 2 &   2.058 &  -0.037 &  -0.593 &  -0.006 &   1.423 \\
              & 3 & 3 &   1.643 &  -0.036 &  -0.530 &   0.006 &   1.084 \\
              & 4 & 3 &  -2.500 &  -0.039 &   0.000 &   0.148 &  -2.391 \\
\end{tabular}
\end{ruledtabular}
\end{table}

\section{Results}

The results of the calculations for the spin-independent part of the
PNC amplitudes ($z$-components) are 
\begin{eqnarray}
{\rm Ba^+:} && E1^{\rm PNC}(5d_{3/2} - 6s) = \nonumber \\
&&0.29(2) \times 10^{-12} Q_W iea_0, \label{eq:bapnc} \\
{\rm Yb^+:} && E1^{\rm PNC}(5d_{3/2} - 6s) = \nonumber \\
&&0.62(20) \times 10^{-12} Q_W iea_0, \label{eq:ybpnc} \\
{\rm Ra^+:} && E1^{\rm PNC}(6d_{3/2} - 7s) = \nonumber \\
&&3.4(1) \times 10^{-12} Q_W iea_0. \label{eq:rapnc} 
\end{eqnarray}
The uncertainties are estimated by comparing {\em ab initio}
calculations with the calculations in which matrix elements were
rescaled as it was described in previous section.
The expressions (\ref{eq:bapnc},\ref{eq:ybpnc},\ref{eq:rapnc}) are
valid for any isotopes. All dependence on nuclear number $A$ is via
weak nuclear charge $Q_W$ (see, (\ref{eq:qw})) while dependence on
nuclear radius is negligible. To be precise, the dependence of the PNC
amplitudes on the nuclear radius can be included with the help of an
additional factor
\begin{equation}
  E1^{\rm PNC}(A_2) =
  \left(\frac{A_2}{A_1}\right)^{-\frac{Z^2\alpha^2}{3}} E1^{\rm PNC}(A_1).
\label{eq:pnca}
\end{equation}
For cases considered in this work the maximum value of the correction
is 0.4\% (between $^{223}$Ra and $^{229}$Ra). For other cases the
correction is even smaller. This is beyond the accuracy of present
calculations. 

It is convenient to present the total PNC amplitude (including the
spin-dependent part) in a form
\begin{equation}
  E1^{\rm PNC} = P(1 + R\varkappa),
\label{eq:PR}
\end{equation}
where $P$ is the spin-independent part (including weak nuclear charge
$Q_W$) and $R$ is the ratio of the spin-dependent to the
spin-independent amplitudes. This has two important
advantages~\cite{DF11}: (i) extraction of the value of $\varkappa$
from experimental data can lead to no confusion over its sign, (ii)
the uncertainty for the value of the ratio of the spin-dependent and
spin-independent amplitudes $R$ is usually lower than for each of the
amplitudes. This is because the two amplitudes are very similar
and numerical uncertainty cancels out in the ratio (see also
Ref.~\cite{DF11}). 

The total PNC amplitudes for different hfs transitions in Ba$^+$,
Yb$^+$ and Ra$^+$ are presented in Table~\ref{t:sdpnc}. The table
includes all stable isotopes of Ba and Yb which have non-zero nuclear
spin and the most stable isotopes of Ra with non-zero nuclear spin.
The results for other isotopes can be obtained by rescaling 
appropriate PNC amplitude (with required values of $F_1,F_2$ and $I$)
using corresponding weak nuclear charges: 
\begin{eqnarray}
  &&E1^{\rm PNC}(A_2)_{F_1F_2I} = \label{eq:PRA2} \\
  &&P(A_1)_{F_1F_2I}\frac{Q_W(A_2)}{Q_W(A_1)}\left[1 +
    R(A_1)_{F_1F_2I}\frac{Q_W(A_1)}{Q_W(A_2)}\varkappa\right],   
\nonumber
\end{eqnarray}
where $P(A_1)_{F_1F_2I}$ and $R(A_1)_{F_1F_2I}$ are taken from
Table~\ref{t:sdpnc} and $Q_W(A_1)$ and $Q_W(A_2)$ are calculated using
(\ref{eq:qw}). 
We stress ones more that the dependence of the amplitudes on the
nuclear radius is much smaller than current theoretical uncertainty.

Numerical uncertainties for $P$ and $R$ are presented in parentheses 
in Table~\ref{t:sdpnc}. One can see that for some hyperfine
transitions the uncertainty for $R$ is very low. Comparing the data
in Tables \ref{t:sdpnc} and \ref{t:ssss} reveals that low uncertainty
in $R$ corresponds to the cases when the spin-dependent PNC amplitude
is strongly dominated by the sum $S_1+S_4$ while the sum of two other
terms ($S_2$ and $S_3$) is small. This is because strong domination of
$S_1+S_4$ makes the spin-dependent PNC amplitude to be very similar to
the spin-independent one where $S_2 \equiv 0$ and $S_3 \equiv 0$. In
this case the rescaling changes both amplitudes at the same rate and
the change cancels out in the ratio $R$. The hfs transitions with low
uncertainty in $R$ are good candidates for the measurements when the
aim is extraction of $\varkappa$.

\begin{table}
\caption{PNC amplitudes ($z$-components) for the $|5d_{3/2},F_1
  \rangle \rightarrow  |6s_{1/2},F_2\rangle$ transitions in
  $^{135}$Ba$^+$, $^{137}$Ba$^+$, $^{171}$Yb$^+$ and $^{173}$Yb$^+$  and
$|6d_{3/2},F_1  \rangle \rightarrow  |7s_{1/2},F_2\rangle$ transitions in
  $^{223}$Ra$^+$, $^{225}$Ra$^+$ and $^{229}$Ra$^+$. 
Units: $10^{-10} iea_0$.} 
\label{t:sdpnc}
\begin{ruledtabular}
\begin{tabular}{cccccrcl}
Ion & $Q_W$ & $I$ & $F_1$ & $F_2$ & \multicolumn{3}{c}{PNC amplitude} \\
\hline
$^{135}$Ba$^+$ & -74.11 & 1.5 & 0 & 1 &  $-0.152(9)$ &$\times$ & $[1+0.0409(2)\varkappa]$ \\
              &        &     & 1 & 1 &  $-0.170(11)$ &$\times$ & $[1+0.0400(2)\varkappa]$ \\
              &        &     & 1 & 2 &  $-0.059(4)$ &$\times$ & $[1-0.021(2)\varkappa]$ \\
              &        &     & 2 & 1 &  $0.132(9)$ &$\times$ & $[1+0.039(1)\varkappa]$ \\
              &        &     & 2 & 2 &  $-0.152(9)$ &$\times$ & $[1-0.023(1)\varkappa]$ \\
              &        &     & 3 & 2 &  $0.152(9)$ &$\times$ & $[1-0.0245(1)\varkappa]$ \\

$^{137}$Ba$^+$ & -76.09 & 1.5 & 0 & 1 &  $-0.156(10)$ &$\times$ & $[1+0.0398(2)\varkappa]$ \\
               &        &    & 1 & 1 &  $-0.175(11)$ &$\times$ & $[1+0.0392(3)\varkappa]$ \\
               &        &    & 1 & 2 &  $-0.061(4)$ &$\times$ & $[1-0.021(2)\varkappa]$ \\
               &        &    & 2 & 1 &  $0.135(8)$ &$\times$ & $[1+0.038(1)\varkappa]$ \\
               &        &    & 2 & 2 &  $-0.156(10)$ &$\times$ & $[1-0.022(1)\varkappa]$ \\
               &        &    & 3 & 2 &  $0.156(10)$ &$\times$ & $[1-0.0239(1)\varkappa]$ \\

$^{171}$Yb$^+$ & -94.86 & 0.5 & 1 & 0 &  $~~0.59(19)$ &$\times$ & $[1+0.030(16)\varkappa]$ \\
              &         &    & 1 & 1 &  $-0.29(9)$ &$\times$ & $[1+0.019(2)\varkappa]$ \\
              &         &    & 2 & 1 &  $~~0.51(16)$ &$\times$ & $[1-0.016(6)\varkappa]$ \\
         
$^{173}$Yb$^+$ & -96.84 & 2.5 & 1 & 2 &  $-0.41(13)$ &$\times$ & $[1+0.022(9)\varkappa]$ \\
              &         &    & 2 & 2 &  $-0.53(17)$ &$\times$ & $[1+0.015(8)\varkappa]$ \\
              &         &    & 2 & 3 &  $-0.17(6)$ &$\times$ & $[1+0.009(2)\varkappa]$ \\
              &         &    & 3 & 2 &  $~~0.28(9)$ &$\times$ & $[1+0.005(3)\varkappa]$ \\
              &         &    & 3 & 3 &  $-0.48(5)$ &$\times$ & $[1-0.002(1)\varkappa]$ \\
              &         &    & 4 & 3 &  $~~0.37(12)$ &$\times$ & $[1-0.016(5)\varkappa]$ \\

$^{223}$Ra$^+$ & -127.2 & 1.5 & 0 & 1 &  $-3.04(9)$ &$\times$ & $[1+0.0252(1)\varkappa]$ \\
              &        &     & 1 & 1 &  $-3.40(10)$ &$\times$ & $[1+0.0233(3)\varkappa]$ \\
              &        &     & 1 & 2 &  $-1.18(4)$ &$\times$ & $[1-0.0053(5)\varkappa]$ \\
              &        &     & 2 & 1 &  $2.64(8)$ &$\times$ & $[1+0.0193(3)\varkappa]$ \\
              &        &     & 2 & 2 &  $-3.04(9)$ &$\times$ & $[1-0.0093(4)\varkappa]$ \\
              &        &     & 3 & 2 &  $3.04(9)$ &$\times$ & $[1-0.0151(1)\varkappa]$ \\

$^{225}$Ra$^+$ & -129.2 & 0.5 & 1 & 0 &  $~~4.37(13)$ &$\times$ & $[1+0.0389(3)\varkappa]$ \\
              &        &     & 1 & 1 &  $-2.19(6)$ &$\times$ & $[1-0.0033(6)\varkappa]$ \\
              &        &     & 2 & 1 &  $~~3.79(11)$ &$\times$ & $[1-0.0149(1)\varkappa]$ \\

$^{229}$Ra$^+$ & -133.1 & 2.5 & 1 & 2 &  $-3.02(9)$ &$\times$ & $[1+0.0202(1)\varkappa]$ \\
              &        &     & 2 & 2 &  $-3.97(12)$ &$\times$ & $[1+0.0180(1)\varkappa]$ \\
              &        &     & 2 & 3 &  $-1.27(4)$ &$\times$ & $[1-0.0066(4)\varkappa]$ \\
              &        &     & 3 & 2 &  $~~2.12(6)$ &$\times$ & $[1+0.0146(3)\varkappa]$ \\
              &        &     & 3 & 3 &  $-3.56(10)$ &$\times$ & $[1-0.0100(3)\varkappa]$ \\
              &        &     & 4 & 3 &  $~~2.76(8)$ &$\times$ & $[1-0.0145(1)\varkappa]$ \\
\end{tabular}
\end{ruledtabular}
\end{table}

%Factors for Ba:
%PNC:  0.9868  0.9532
%SD:   0.9868  0.9532  0.9189  1.0237
%E1:   1.0120  1.1493  0.9701  0.9477

%Factors for Ra:
%PNC:  0.9674  0.9214
%SD:   0.9674  0.9214  0.8776  1.0157
%E1:   1.0043  1.0013  1.0047  1.0017

\subsection{Comparison with other calculations}

Table~\ref{t:pnc} summarizes present and past calculations of the
spin-independent PNC s-d amplitudes in Ba$^+$ and Ra$^+$. We present
the results in a form of the coefficients before weak nuclear charge
$Q_W$. These coefficients are practically isotope-independent. This is
because the isotope-dependence of the PNC amplitudes is strongly
dominated by weak nuclear charge while the dependence of the PNC
amplitudes on the details of nuclear density is very weak and can be
neglected on the present level of accuracy.

The technique used in the present work is very similar to the
sum-over-states  approach of our previous paper~\cite{DFG01}. As
expected, the results are very close too. There is also good
agreement with Sahoo {\em et al} for Ba$^+$~\cite{Sahoo06} and with
Wansbeek {\em et al} for Ra$^+$~\cite{Wansbeek} and with recent
calculations by Pal {\em et al}~\cite{Pal09} for Ra$^+$.

\begin{table}
\caption{Spin-independent part of the parity-nonconserving s-d
  amplitudes in Ba$^+$, Yb$^+$ and Ra$^+$. Units: $10^{-12} Q_W iea_0$.}
\label{t:pnc}
\begin{ruledtabular}
\begin{tabular}{ccll}
Ion & Transition & \multicolumn{1}{c}{This work} &\multicolumn{1}{c}{Other} \\
\hline
Ba$^+$ & $5d_{3/2} - 6s_{1/2}$ & 0.29(2) & 0.29\footnotemark[1],\ 0.304\footnotemark[2] \\
Yb$^+$ & $5d_{3/2} - 6s_{1/2}$ & 0.62(20) & - \\
%0.57(11)\footnotemark[2] \\
Ra$^+$ & $6d_{3/2} - 7s_{1/2}$ & 3.4(1) &
3.3\footnotemark[1],\ 3.36\footnotemark[3], \ 3.33\footnotemark[4] \\ 
\end{tabular}
\footnotemark[1]{Ref.~\cite{DFG01}.}
\footnotemark[2]{Ref.~\cite{Sahoo06}.}
%\footnotemark[2]{Porsev.}
\footnotemark[3]{Ref.~\cite{Wansbeek}.}
\footnotemark[4]{Ref.~\cite{Pal09}.}
\end{ruledtabular}
\end{table}

\begin{table}
\caption{Reduced matrix of the spin-dependent parity-nonconserving s-d
  amplitudes in Ba$^+$ and Ra$^+$. Units: $10^{-12} \varkappa iea_0$.}
\label{t:sahoo}
\begin{ruledtabular}
\begin{tabular}{cccccdd}
Ion & Transition & $I$ & $F_1$ & $F_2$ & \multicolumn{1}{c}{\rm This work}
&\multicolumn{1}{c}{\rm Ref.~\cite{Sahoo11}} \\ 
\hline
$^{135}$Ba$^+$ & $6s_{1/2} - 5d_{3/2}$ & 1.5 & 2 & 3 & -1.71 & -1.94 \\
              &                      &     & 1 & 2 &  1.62 &  1.79 \\
$^{139}$Ba$^+$ & $6s_{1/2} - 5d_{3/2}$ & 3.5 & 3 & 3 & -1.86 & -2.07 \\
              &                      &     & 3 & 2 &  1.86 &  2.11 \\
\hline
$^{225}$Ra$^+$ & $7s_{1/2} - 6d_{3/2}$ & 0.5 & 1 & 2 & -17.8 & -19.8 \\
$^{223}$Ra$^+$ & $7s_{1/2} - 6d_{3/2}$ & 1.5 & 2 & 3 & -21.1 & -23.5 \\
              &                      &     & 1 & 2 &  16.1 &  20.3 \\
$^{229}$Ra$^+$ & $7s_{1/2} - 6d_{3/2}$ & 2.5 & 2 & 3 & -3.8 & -6.5 \\
              &                      &     & 2 & 2 & -19.6 & -22.9 \\
\end{tabular}
\end{ruledtabular}
\end{table}

Table~\ref{t:sahoo} compares our calculated reduced matrix elements of
the spin-dependent PNC amplitudes with the results of the recent
calculations by Sahoo {\em et al}~\cite{Sahoo11}. To make the
comparison easy we have multiplied all matrix elements from~\cite{Sahoo11}
by 2 and have changed their signs. The former is to take into account
different definition of $\varkappa$, the latter is due to the fact that
we also have an opposite sign for the spin-independent PNC amplitude
compared to what is presented in \cite{Wansbeek} and \cite{Sahoo11}.
The total sign of an amplitude is not fixed and can be changed
arbitrarily. Note however that the relative sign of the SI and SD PNC
amplitudes is not arbitrary and the sign can only be changed for both
parts of the amplitudes simultaneously.

Given that the accuracy of the present calculations is few per cents and
similar accuracy should be expected for \cite{Sahoo11} the results
presented in Table~\ref{t:sahoo} are in a reasonable agreement with
each other. Comparison of the
data in Table~\ref{t:sahoo} and Table~\ref{t:ssss} shows that the
difference between our results and those of Sahoo {\em et al} is
larger for cases when there is strong cancelation between the $S_1,
S_2,  S_3$ and $S_4$ contributions to the reduced matrix element. For
example, the largest difference is for the $F_1=2$ to $F_2=3$ 
transition in $^{229}$Ra$^+$. The data in Table~\ref{t:ssss} shows
that the final value of the reduced matrix element for this case is
just about 40\% of the $S_1$ contribution. On the
contrary, if the amplitude is dominated by the $S_1$ term the agreement
between results of the two works is much better. This should be
expected since the $S_1$ term is the most stable in the
calculations. 
%Therefore, the differences in the values of matrix
%elements of the two works is likely to be a pure numerical problem
%cased by insufficient accuracy of calculations. 

\section{Conclusion}

We present simultaneous calculation of the spin-independent and
spin-dependent PNC amplitudes of the s-d transitions in Ba$^+$, Yb$^+$
and Ra$^+$. The results are to be used for
accurate interpretation of future measurements in terms of the
parameter of the spin-dependent PNC interaction $\varkappa$. Both,
sign and value of $\varkappa$ can be determined. Theoretical
uncertainty is at the level of 3 to 6\% for Ba$^+$ and Ra$^+$ and 30
to 50\% for Yb$^+$. Note that the uncertainty for the spin-independent
PNC amplitude can be further reduced by including structure radiation
and ladder diagrams for more accurate treatment of correlations and by
including other small corrections (Breit, QED, etc.).
The uncertainty for the relative contribution of the
nuclear-spin-dependent part of the PNC amplitude is already small 
being on the level of 1\% in some cases.
The ratio of the SD to SI PNC amplitude is to be measured to extract
the calue of $\varkappa$.
The results of the PNC calculations for Yb$^+$ are presented for the
first time.

\acknowledgements

The authors are grateful to M. G. Kozlov and S. G. Porsev for useful
discussions. 
The work was supported in part by the Australian Research Council.

\end{document}